\documentclass{llncs}

\usepackage{graphicx}
\usepackage{fancyvrb}
\usepackage{verbatim}
\usepackage{color}
\usepackage{array}
\usepackage{subcaption}
\fvset{frame=single,samepage=true,fontfamily=courier,numbers=left,numbersep=2pt,baselinestretch=0.90,fontsize=\scriptsize,commandchars=\\\{\}}

\begin{document}

\title{Support for Various HTTP Methods on the Web}
\author{Sawood Alam and Charles L. Cartledge and Michael L. Nelson\\
    \email{\{salam,ccartled,mln\}@cs.odu.edu}
    \institute{Department of Computer Science\\
        Old Dominion University\\
        Norfolk, VA 23529 USA
    }
}

\maketitle

\begin{abstract}
We examine how well various HTTP methods are supported by public web services. We sample 40,870 live URIs from the DMOZ collection (a curated directory of World Wide Web URIs) and found that about 55\% URIs claim support (in the Allow header) for GET and POST methods, but less than 2\% of the URIs claim support for one or more of PUT, PATCH, or DELETE methods.
\end{abstract}

\section{Introduction}

Tim Berners-Lee invented the World Wide Web (WWW) in 1989~\cite{webhist,tbcppf,tbclcstalk,tbcfpw}. Since then the Web has evolved in many phases. There is no agreed upon well-defined name for these phases, but for the sake of understanding some people have divided these phases as follows~\cite{webbird,defweb,embraceweb,emergeweb,web1to4,web1to5}:

\begin{description}
\item[Web 0.0:] This was the pre-web era when the Internet was developing.
\item[Web 1.0:] The initial phase of the Web is called ``Read-Only Web''. Since the invention of the Web until 1999 there were mostly static HTML pages connected with hyperlinks. Very few people were responsible for putting information online that could be searched and consumed by others. There was very limited flow of information from the user to the server. This phase is also called as ``shopping cart and static web'' era.
\item[Web 2.0:] The second major change in the Web came when collaborative applications like wikis, self-publishing platforms like blogs, and social media like Twitter, Flickr, and YouTube emerged. This phase is called ``Read-Write Web'' or ``Read-Write-Publish Era''. In this phase the consumer of the content became first class producer of the content as well. This increased participation from the Internet user increased the size of the Web and reduced the gap between the number of producers and the number of consumer of the information on the Web.
\item[Web 3.0:] The third major change in the Web began after the inclusion of semantic markup. This phase is called ``Read-Write-Execute Web'', ``Semantic Web'', or ``Semantic Executing Web''. Semantic markups allow machines to understand and link information available on the Web. Usage of Resource Description Framework (RDF)~\cite{rdf} and Web Ontology Language (OWL)~\cite{owl} makes the machines as the first class consumer of the information on the Web.
\item[Web 4.0:] This is the upcoming Web or the next Web. It is still in the very early stage of taking its shape, hence it is not very clear what it will eventually look like. It is called ``Read-Write-Execution-Concurrency Web'' or ``Open, Linked, and Intelligent Web''. In this phase of the Web, machines will not only be able to consume and understand linked semantic data but will also be able to produce information based on their learning. Machines will be able to communicate with other machines and resources like we humans do. Because of this nature, it is also called ``Symbiotic Web''.
\item[Web 5.0:] This phase is an idea in progress and definition is yet unknown. It is called ``Symbionet Web'' which will be very decentralized in nature in which devices/machines will be able to explore other interconnected devices and create the model of the Web. It will also have the component of emotion while the current Web is emotionally neutral.
\end{description}

While the Web was evolving, the underlying web communication protocol known as Hypertext Transfer Protocol (HTTP)~\cite{rfc2616} was also evolving to fulfil the increased communication needs. Initially the HTTP methods defined in the protocol were not fully implemented by web servers. As the Web graduated from Web 1.0 to 2.0 and beyond, the need for a web-scale distributed software architecture increased which gave birth to REST (REpresentational State Transfer)~\cite{fielding2000architectural,fielding2002principled}. HTTP and its extensions define a rich set of methods to interact with web resources. To perform CRUD (Create, Read, Update, and Delete) operations on a resource, commonly used HTTP methods are GET, POST, PUT, PATCH, and DELETE. Table~\ref{tab:methodsactions} summarizes various HTTP methods and their corresponding resource actions. Utilizing these HTTP methods for appropriate resource actions along with media types and relation types in an HTTP communication is called REST.

\begin{table}[!t]
  \centering
  \caption{HTTP Method Mapping with Resource Action}
  \label{tab:methodsactions}
  \begin{tabular}{p{2.5cm}|p{9.0cm}}
    \hline
    \textbf{HTTP Method} & \textbf{Resource Action}\\
    \hline
    GET & Retrieve representation of a resource.\\
    POST & Create a new resource.\\
    PUT & Overwrite an existing resource or create new if it does not exist.\\
    PATCH & Partially update a resource.\\
    DELETE & Delete a resource.\\
    \hline
  \end{tabular}
\end{table}

While REST utilizes various HTTP methods for various resource actions, Remote Procedure Call (RPC) encourages application designers to define their own application specific methods and rely only on GET and POST HTTP methods. Limiting a web service to only GET and POST methods has some consequences that lead to non-standard usage of these methods. For example, POST is a non-idempotent method, if it is used to update a resource in place of PUT, which is an idempotent method, then it will violate the definition of the POST method. An idempotent method is a method in which multiple requests of the same method have the same effect on the resource as if the request was made just once. Another consequence is related to the cacheability of methods. For instance, response to a GET request is cacheable, if GET method is used in place of a non-cacheable method like DELETE, it may result in undesired effects. Table~\ref{tab:restrpc} compares REST and RPC style URIs to illustrate utilization of HTTP methods in REST and application specific methods in RPC.

\begin{table}[!t]
  \centering
  \caption{Example URI Constructs in REST vs. RPC}
  \label{tab:restrpc}
  \begin{tabular}{p{3.2cm}|p{5.2cm}}
    \hline
    \textbf{REST} & \textbf{RPC}\\
    \hline
    GET \hspace{2.0em} /tasks   & GET \hspace{0.7em} /list\_all\_tasks.php\\
    GET \hspace{2.0em} /tasks/1 & GET \hspace{0.7em} /show\_task\_details.php?id=1\\
    POST \hspace{1.5em} /tasks   & POST \hspace{0.15em} /create\_new\_task.php\\
    PATCH \hspace{0.8em} /tasks/1 & POST \hspace{0.15em} /update\_task\_status.php\\
    DELETE \hspace{0.15em} /tasks/1 & GET \hspace{0.7em} /delete\_task.php?id=1\\
    \hline
  \end{tabular}
\end{table}

Unfortunately, many web services do not support all of the HTTP methods. Hence, a PATCH request (or other methods like PUT or DELETE) may cause the server to respond with \texttt{501 Not Implemented} or other failure responses. For example, the default Apache web server setup returns \texttt{405 Method Not Allowed} in response to a PUT request.

Table~\ref{tab:methods} lists common HTTP methods and their support in web browsers and the most common LAMP\footnote[1]{Linux, Apache, MySQL, and PHP, Perl or Python.} servers stack. It shows that Apache web server requires extra configuration in order to support PUT, DELETE and PATCH methods. Also, pure HTML has no interface to issue these methods from the browser except by using Ajax\footnote[2]{Asynchronous JavaScript and XML} requests.

We explore the support of HTTP methods on the live Web, that are responsible for RESTful HTTP communication. Previously, we performed a brief analysis of HTTP method support to show the limited support of many HTTP methods~\cite{hmtech}. In this paper, we examine server response headers on live URIs sampled from the DMOZ~\cite{dmoz}. The DMOZ is an Open Directory Project that maintains the curated directory of World Wide Web URIs, currently listing over 5 million URIs.

\begin{table}[!t]
  \centering
  \caption{HTTP Method Support}
  \label{tab:methods}
  \begin{tabular}{p{2.2cm}|p{2.8cm} p{2.2cm} p{1.0cm}}
    \hline
    \textbf{{\color{black}Method}} & \textbf{{\color{black}LAMP}} & \textbf{{\color{black}HTML}} & \textbf{{\color{black}Ajax}}\\
    \hline
    {\color{black}GET} & {\color{blue}Default Support} & {\color{blue}Link, Form} & {\color{blue}Yes}\\[-0.25cm]
    {\color{black}POST} & {\color{blue}Default Support} & {\color{blue}Form} & {\color{blue}Yes}\\[-0.25cm]
    {\color{black}HEAD} & {\color{blue}Default Support} & {\color{red}None} & {\color{blue}Yes}\\[-0.25cm]
    {\color{black}OPTIONS} & {\color{blue}Default Support} & {\color{red}None} & {\color{blue}Yes}\\[-0.25cm]
    {\color{black}PUT} & {\color{red}Extra Config.} & {\color{red}None} & {\color{blue}Yes}\\[-0.25cm]
    {\color{black}DELETE} & {\color{red}Extra Config.} & {\color{red}None} & {\color{blue}Yes}\\[-0.25cm]
    {\color{black}PATCH} & {\color{red}Extra Config.} & {\color{red}None} & {\color{blue}Yes}\\
    \hline
  \end{tabular}
\end{table}

\section{Background}

The REST application architecture allows clients to interact with any RESTful service without any out-of-band information. To facilitate this feature, REST applies the principle of Hypermedia as the Engine of Application State (HATEOAS)~\cite{fielding2002principled}. According to this principle, the interaction with a REST service begins from a well-known URI and from there the REST service guides the clients with the help of links corresponding to state transitions (the same way people follow links in the HTML pages). A client with generic understanding of relation types and media types can easily follow the options towards its interaction path. There are two ways for a client to discover the capabilities of a server for a given resource URI: the OPTIONS method and the ``Allow'' header.

The OPTIONS method is utilized to discover the capabilities of the server and specific HTTP methods supported for the given URI. Figure~\ref{img:optallow} illustrates an OPTIONS request and corresponding response from the server.

\begin{figure}
\begin{Verbatim}
$ curl -I \textcolor{red}{-X OPTIONS} http://www.cs.odu.edu/
HTTP/1.1 200 OK
Date: Wed, 07 Aug 2013 23:11:04 GMT
Server: Apache/2.2.17 (Unix) PHP/5.3.5 mod_ssl/2.2.17 OpenSSL/0.9.8q
\textcolor{red}{Allow: GET,HEAD,POST,OPTIONS}
Content-Length: 0
Content-Type: text/html

$
\end{Verbatim}
\caption{OPTIONS Request and Response}
\label{img:optallow}
\end{figure}

The ``Allow'' header is an entity header that gives the comma-separated list of allowed methods on a given resource URI. This header should be included in the response to an OPTIONS request. According to the HTTP specification, the ``Allow'' header must be present if the server returns a \texttt{405 (Method Not Allowed)} response. Line 5 of Figure~\ref{img:optallow} illustrates an ``Allow'' header which tells the client that only GET, HEAD, POST, and OPTIONS methods are supported for this URI.

\section{Methodology}

In order to find the supported methods on URIs we issued OPTIONS request on them. If a web server implements OPTIONS method properly, it returns ``Allow'' header in the response to the OPTIONS request and lists the supported methods on the given URI.

\begin{figure}[!t]
\centering
\includegraphics[width=0.8\linewidth]{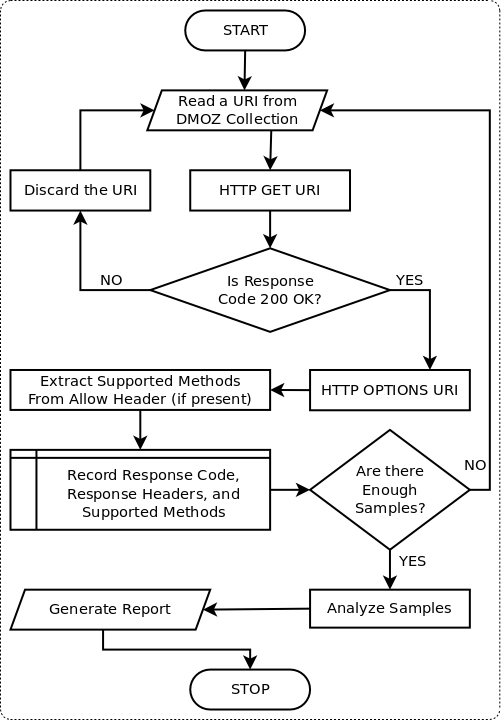}
\caption{Process of Method Support Analysis.}
\label{img:optanalysis}
\end{figure}

To measure the distribution of HTTP method support on the Web, we randomly chose 100,000 HTTP URIs from a historical collection of the DMOZ that was created for a prior study~\cite{howmuchwebarch}. These URIs include both live and dead URIs. Initial sampling of the URIs was done without considering any filtering or enforced distribution of Top Level Domains (TLDs) or the path depth of the URIs. Figure~\ref{img:optanalysis} illustrates the process utilized in the analysis of support of HTTP methods on the Web.

\begin{table}[!ht]
  \centering
  \caption{Path Depths for URIs in Sample Set}
  \label{tab:uridepth}
  \begin{tabular}{p{2.2cm} p{5.6cm} | >{\raggedleft\arraybackslash}p{2cm} >{\raggedleft\arraybackslash}p{2cm}}
    \hline
    \textbf{{\color{black}Path Depth}} & \textbf{{\color{black}Example}} & \textbf{{\color{black}Occurrence}} & \textbf{{\color{black}Percentage}}\\
    \hline
    0 & example.com & 27718 & 67.820\%\\
    1 & example.com/a & 5141 & 12.579\%\\
    2 & example.com/a/b & 4616 & 11.294\%\\
    3 & example.com/a/b/c & 1689 & 4.133\%\\
    4 & example.com/a/b/c/d & 765 & 1.872\%\\
    5 & example.com/a/b/c/d/e & 405 & 0.991\%\\
    6 & example.com/a/b/c/d/e/f & 373 & 0.913\%\\
    7 & example.com/a/b/c/d/e/f/g & 128 & 0.313\%\\
    8 & example.com/a/b/c/d/e/f/g/h & 29 & 0.071\%\\
    9 & example.com/a/b/c/d/e/f/g/h/i & 3 & 0.007\%\\
    10 & example.com/a/b/c/d/e/f/g/h/i/j & 1 & 0.002\%\\
    11 & example.com/a/b/c/d/e/f/g/h/i/j/k & 1 & 0.002\%\\
    12 & example.com/a/b/c/d/e/f/g/h/i/j/k/l & 1 & 0.002\%\\
    \hline
  \end{tabular}
\end{table}

\begin{figure}[!t]
\includegraphics[width=\linewidth]{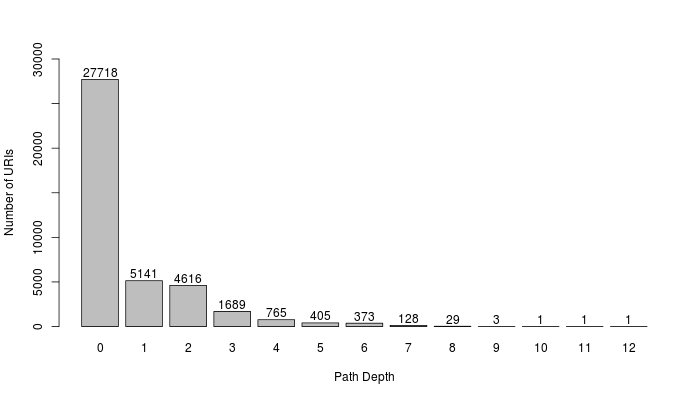}
\caption{Path Depth for URIs in Sample Set.}
\label{img:pdepth}
\end{figure}

\begin{table}[t]
  \centering
  \caption{TLD Distribution for Sample Set URIs}
  \label{tab:uritld}
  \begin{tabular}{p{2.5cm} | >{\raggedleft\arraybackslash}p{2.4cm} >{\raggedleft\arraybackslash}p{2.4cm}}
    \hline
    \textbf{{\color{black}TLD}} & \textbf{{\color{black}Occurrence}} & \textbf{{\color{black}Percentage}}\\
    \hline
    .com & 18089 & 44.260\%\\
    .org & 3592 & 8.789\%\\
    .net & 2039 & 4.988\%\\
    .edu* & 875 & 2.141\%\\
    .gov* & 358 & 0.876\%\\
    Others & 15917 & 38.945\%\\
    \hline
  \end{tabular}
\end{table}

According to the HTTP specification, all general-purpose servers must support GET and HEAD methods. In practice though, HEAD is not always supported. Hence for further analysis, we decided to filter only URIs that return a \texttt{200 OK} response to a GET request and we ignored all the other URIs. Out of those 100,000 URIs, there were only 40,870 URIs responding with a 200 status code. Other URIs returned \texttt{3xx}, \texttt{4xx}, \texttt{5xx} status codes, request timeout, or Domain Name Service (DNS) errors. Table~\ref{tab:uritld} shows the distribution of live URIs over various TLDs while Table~\ref{tab:uridepth} and Figure~\ref{img:pdepth} show the distribution of path depths in the sample set. The path depth here refers to the number of forward slashes (\texttt{/}) excluding the trailing slash, if any. Also, the path depth does not consider hash portion of the URI or any query parameters. These 40,870 live URIs belong to 33,183 different domains. We issued an OPTIONS request on each of those 40,870 live URIs to collect data about supported methods from the ``Allow'' response header as illustrated in Figure~\ref{img:optallow}. Table~\ref{tab:optstatus} shows the distribution of response codes for OPTIONS requests over 40,870 URIs. We did not follow the 58 (0.142\%) URIs returning a redirection status code. We do not feel that these URIs would have any significant impact on the results.

\begin{table}[t]
  \centering
  \caption{Response Codes for OPTIONS Requests}
  \label{tab:optstatus}
  \begin{tabular}{p{2.6cm} | >{\raggedleft\arraybackslash}p{2.6cm} >{\raggedleft\arraybackslash}p{2.4cm}}
    \hline
    \textbf{{\color{black}Status Code}} & \textbf{{\color{black}Occurrence}} & \textbf{{\color{black}Percentage}}\\
    \hline
    \textbf{200} & 37347 & 91.380\% \\
    \hline
    301 & 6 & \\
    302 & 50 & \\
    303 & 2 & \\
    \textbf{3xx} & 58 & 0.142\% \\
    \hline
    400 & 44 & \\
    401 & 30 & \\
    403 & 1112 & \\
    404 & 326 & \\
    405 & 1312 & \\
    406 & 7 & \\
    407 & 1 & \\
    411 & 3 & \\
    417 & 1 & \\
    422 & 1 & \\
    \textbf{4xx} & 2837 & 6.942\% \\
    \hline
    500 & 100 & \\
    501 & 479 & \\
    502 & 34 & \\
    503 & 14 & \\
    \textbf{5xx} & 627 & 1.534\% \\
    \hline
    \textbf{999} & 1 & 0.002\% \\
    \hline
    \textbf{Total} & 40870 & \\
    \hline
  \end{tabular}
\end{table}

\section{Scenarios}

We will look at samples of four different OPTIONS requests with different responses. The first request as illustrated in Figure~\ref{img:opt501} returned \texttt{501 Not Implemented} response, which means it does not recognize OPTIONS method, although the URI associated with this request supports POST, GET, HEAD, PUT, and DELETE methods (according to its documentation,) but there is no RESTful way to discover the server capabilities.

The second request as illustrated in Figure~\ref{img:opt405} does recognize the OPTIONS method, but returned \texttt{405 Not Allowed} response. According to the HTTP specification, in case of 405 response, an ``Allow'' header must be present in the response, but in this second request the server does not honor this rule, hence we cannot query supported methods.

The third request as illustrated in Figure~\ref{img:optlimit} returned limited method support (only GET and POST methods listed in Table~\ref{tab:methods}).

Finally, the fourth request as illustrated in Figure~\ref{img:optfull} returned support for all the methods listed in Table~\ref{tab:methods}. We did not check to see if the URIs respond to the methods returned in the ``Allow'' header.

\begin{figure}[!h]
\begin{minipage}[b]{\linewidth}
\centering
\begin{Verbatim}
$ curl -I -X OPTIONS http://fluiddb.fluidinfo.com/about
HTTP/1.1 \textcolor{red}{501 Not Implemented}
Server: nginx/1.1.19
Date: Wed, 07 Aug 2013 21:09:15 GMT
Content-Type: text/html; charset=utf-8
Content-Length: 150
Connection: keep-alive
X-Fluiddb-Request-Id: API-9006-20130807-210915-18792385
X-Fluiddb-Error-Class: UnsupportedMethod

$
\end{Verbatim}
\vspace*{-5mm}
\subcaption{Method Not Implemented.}\label{img:opt501}
\end{minipage}
\begin{minipage}[b]{\linewidth}
\centering
\vspace*{3mm}
\begin{Verbatim}
$ curl -I -X OPTIONS http://dev.bitly.com/
HTTP/1.1 \textcolor{red}{405 Not Allowed}
Content-Type: text/html
Date: Wed, 07 Aug 2013 22:24:05 GMT
Server: nginx
Content-Length: 166
Connection: keep-alive

$
\end{Verbatim}
\vspace*{-5mm}
\subcaption{Method Not Allowed.}\label{img:opt405}
\end{minipage}
\begin{minipage}[b]{\linewidth}
\centering
\vspace*{3mm}
\begin{Verbatim}
$ curl -I -X OPTIONS http://www.cs.odu.edu/
HTTP/1.1 200 OK
Date: Wed, 07 Aug 2013 23:11:04 GMT
Server: Apache/2.2.17 (Unix) PHP/5.3.5 mod_ssl/2.2.17 OpenSSL/0.9.8q
Allow: \textcolor{red}{GET,HEAD,POST,OPTIONS}
Content-Length: 0
Content-Type: text/html

$
\end{Verbatim}
\vspace*{-5mm}
\subcaption{Limited CRUD Support.}\label{img:optlimit}
\end{minipage}
\begin{minipage}[b]{\linewidth}
\centering
\vspace*{3mm}
\begin{Verbatim}
$ curl -I -X OPTIONS http://www.parasitesandvectors.com/
HTTP/1.1 200 OK
Set-Cookie: UUID=6818dd14-085e-4a50-806a-deab9b907585; Path=/
Allow: \textcolor{red}{GET, HEAD, POST, PUT, DELETE, TRACE, OPTIONS, PATCH}
Content-Type: text/html
X-Cacheable: NO
Server: BioMed Central Web Server 1.0
Content-Length: 0
Accept-Ranges: bytes
Date: Wed, 07 Aug 2013 23:02:22 GMT
Connection: keep-alive

$
\end{Verbatim}
\vspace*{-5mm}
\subcaption{Full CRUD Support.}\label{img:optfull}
\end{minipage}
\caption{OPTIONS Method Scenarios.}\label{img:scenario}
\end{figure}

\section{Response Issues}

During the analysis of the response we found various issues including incorrect implementation, malformed headers, and lack of compliance with the standards.

\subsection{Client-dependent Response}

In our experiment, we found that some servers change their response based on the ``User-Agent'' string passed in the request. Figure~\ref{img:useragent} illustrates such a case where if we do not pass a ``User-Agent'' it returns response code \texttt{405}, but if we pass ``Mozilla'' or some other common user-agents, the response code changes to \texttt{200} along with changed response headers.

\subsection{Inaccuracies}

Figure~\ref{img:useragent} illustrates a discrepancy in the response where Line~2 says, \texttt{405 Method Not Allowed}, but in the Line~3 OPTIONS method is listed in the ``Allow'' header.

Further down in Figure~\ref{img:useragent} at Line 12 it says, \texttt{200 OK}, but there is no ``Allow'' header in the response headers below.

In Figure~\ref{img:opt405} the response code is \texttt{405 Not Allowed}, hence it must tell what methods are allowed, but it does not return ``Allow'' header in the response.

\begin{figure}[h]
\begin{Verbatim}
$ curl -I -X OPTIONS http://www.romebedbreakfast.it/
HTTP/1.1 \textcolor{red}{405 Method Not Allowed}
Allow: \textcolor{red}{GET, HEAD, OPTIONS, TRACE}
Content-Type: text/html
Server: Microsoft-IIS/7.5
X-Powered-By: ASP.NET
X-Powered-By-Plesk: PleskWin
Date: Sun, 28 Apr 2013 23:50:39 GMT
Content-Length: 1293

$ curl -I -X OPTIONS \textcolor{blue}{-A "Mozilla"} http://www.romebedbreakfast.it/
HTTP/1.1 \textcolor{red}{200 OK}
Date: Sun, 28 Apr 2013 23:53:17 GMT
Server: Apache
Vary: Accept-Encoding
X-Powered-By: PleskLin
Cache-Control: max-age=0, no-cache
Content-Length: 0
Connection: close
Content-Type: text/html

$
\end{Verbatim}
%$
\caption{Response Changes for Different User-Agents}
\label{img:useragent}
\end{figure}

\subsection{Malformed Allow Headers}

In our experiments we found various types of malformed ``Allow'' headers. We have categorized these malformed ``Allow'' headers in order to study their severity.

\begin{figure}[!h]
\begin{minipage}[b]{\linewidth}
\centering
\begin{Verbatim}
$ curl -I -X OPTIONS http://shwepla.net/index.htm
HTTP/1.1 200 OK
Date: Tue, 17 Sep 2013 16:35:11 GMT
Server: Apache
Allow: OPTIONS,,HEAD,O\textcolor{red}{[RS]},HEAD,HEAD,,HEAD,,HEAD,\textcolor{red}{[STX]},,HEAD,POST,,HEAD,
 ,HEAD,,HEAD,\textcolor{red}{[STX]},/vhosts/,GET,HEAD
X-Powered-By: PleskLin
Content-Length: 0
Connection: close
Content-Type: text/html
X-Pad: avoid browser bug

$
\end{Verbatim}
\vspace*{-5mm}
\subcaption{Allow Header with Control Characters.}\label{img:malspec}
\end{minipage}
\begin{minipage}[b]{\linewidth}
\centering
\vspace*{3mm}
\begin{Verbatim}
$ curl -I -X OPTIONS http://zpiktures.free.fr/
HTTP/1.1 405 Method Not Allowed
Date: Tue, 17 Sep 2013 17:53:53 GMT
Server: Apache/ProXad [Apr 20 2012 15:06:05]
Connection: close
Allow: GET, HEAD, POST, PUT, \textcolor{red}{DELETEtext/html; charset=iso-8859-1}
Cache-Control: no-cache, no-store, must-revalidate
Content-Type: text/html; charset=iso-8859-1

$
\end{Verbatim}
\vspace*{-5mm}
\subcaption{Allow Header with Malformed Method Names.}\label{img:malnames}
\end{minipage}
\begin{minipage}[b]{\linewidth}
\centering
\vspace*{3mm}
\begin{Verbatim}
$ curl -I -X OPTIONS http://www.atexpc.ro/
HTTP/1.1 405 METHOD NOT ALLOWED
Date: Mon, 29 Apr 2013 04:57:46 GMT
Server: Apache/2.2.16 (Debian)
Allow: \textcolor{red}{get}, \textcolor{red}{head}
Vary: Accept-Encoding
Content-Length: 0
Content-Type: text/html; charset=utf-8

$
\end{Verbatim}
\vspace*{-5mm}
\subcaption{Allow Header with Lower-case Method Names.}\label{img:mallower}
\end{minipage}
\begin{minipage}[b]{\linewidth}
\centering
\vspace*{3mm}
\begin{Verbatim}
$ curl -I -X OPTIONS http://www.pembrokepools.co.uk/
HTTP/1.1 200 OK
Date: Mon, 29 Apr 2013 05:01:19 GMT
Server: ECS (ams/498C)
Allow: \textcolor{red}{GET POST OPTIONS}

$
\end{Verbatim}
\vspace*{-5mm}
\subcaption{Allow Header with Space-separated Method Names.}\label{img:malspace}
\end{minipage}
\caption{Malformed Allow Headers.}\label{img:malhead}
\end{figure}

\subsubsection{Special Characters:}

Figure~\ref{img:malspec} illustrates an ``Allow'' header with control characters in it. In our experiment we observed three URIs with this issue. These ``Allow'' headers have returned various special characters including Start of Text (STX), Unit Separator (US), Record Separator (RS), Shift Out (SO), Cancel Character (CAN), and Device Control 4 (DC4).

\subsubsection{Malformed Method Names:}

Figure~\ref{img:malnames} illustrates an ``Allow'' header which has malformed method name. It looks as if the value of the Content-Type header was concatenated with the value of ``Allow'' header. There were 45 such URIs in our sample collection with this issue. We further investigated and found that they all belong to one domain (but different sub-domains).

\subsubsection{Lower-case Method Names:}

Figure~\ref{img:mallower} illustrates lower-cased method names in the ``Allow'' header which is a violation of the specification. According to the HTTP/1.1 specification, method names are case sensitive and general purpose methods like GET and POST are defined as upper-case tokens in the specification. There was only one occurrence of lower-case method names in our test set.

\subsubsection{Space-separated Method Names:}

The list of methods in the ``Allow'' header should be separated by commas. Figure~\ref{img:malspace} illustrates an instance where the method names are separated by spaces.

\subsubsection{Miscellaneous:}

Other issues that we observed in our test set include repeated methods and blank methods. Figure~\ref{img:malspec} illustrates both of these issues. Although the grammar for ``Allow'' header allows one or more commas and optional linear white space (LWS) as the method name separator, also it does not mention anywhere that the list should have unique values in it, hence at the time of parsing the values one needs to be careful about these.

\section{Results}

\begin{table}[!t]
  \centering
  \caption{Summarized Method Support Distribution}
  \label{tab:methodsum}
  \begin{tabular}{p{2.2cm} p{1.6cm} | >{\raggedleft\arraybackslash}p{3.4cm} >{\raggedleft\arraybackslash}p{2.2cm}}
    \hline
    \textbf{{\color{black}Method}} & \textbf{{\color{black}RFC}} & \textbf{{\color{black}Supported (Count)}} & \textbf{{\color{black}Percentage}}\\
    \hline
    GET & 2616 & 22899 & 56.029\%\\
    HEAD & 2616 & 22879 & 55.980\%\\
    OPTIONS & 2616 & 22726 & 55.606\%\\
    POST & 2616 & 16497 & 40.365\%\\
    TRACE & 2616 & 14946 & 36.570\%\\
    DELETE & 2616 & 735 & 1.798\%\\
    PUT & 2616 & 696 & 1.703\%\\
    CONNECT & 2616 & 422 & 1.033\%\\
    \hline
    PROPFIND & 2518 & 1226 & 3.000\%\\
    COPY & 2518 & 1218 & 2.980\%\\
    LOCK & 2518 & 1196 & 2.926\%\\
    UNLOCK & 2518 & 1190 & 2.912\%\\
    MOVE & 2518 & 542 & 1.326\%\\
    PROPPATCH & 2518 & 536 & 1.311\%\\
    MKCOL & 2518 & 523 & 1.280\%\\
    \hline
    MKDIR & 1813 & 6 & 0.015\%\\
    RMDIR & 1813 & 6 & 0.015\%\\
    \hline
    PATCH & 5789 & 418 & 1.023\%\\
    \hline
    REPORT & 3253 & 1 & 0.002\%\\
    \hline
    ACL & 3744 & 1 & 0.002\%\\
    \hline
    SEARCH & 5323 & 611 & 1.495\%\\
    \hline
    INDEX & Unknown & 6 & 0.015\%\\
    NNOC & Unknown & 1 & 0.002\%\\
    \hline
  \end{tabular}
\end{table}

Table~\ref{tab:methodsum} quantifies support of individual HTTP methods described in RFCs~\cite{rfc2616,rfc2518,rfc1813,rfc5789,rfc3253,rfc3744,rfc5323} over a collection of 40,870 live sample URIs. In Table~\ref{tab:methodgroup} common HTTP methods are grouped together to represent the method support distribution for each category. These categories include:

\begin{description}
\item[Safe:] The safe category refers to the set of methods that are only intended to retrieve information without changing the state of the resource. These safe methods include HEAD, GET, and OPTIONS.
\item[Non-safe:] The non-safe category refers to the set of methods that changes the state of the resource and it includes POST, PUT, PATCH, and DELETE methods.
\item[Idempotent:] The idempotent category refers to the set of methods in which multiple requests of the same method have the same effect on the resource as if the request was made just once. All the safe methods are idempotent but some non-safe methods are idempotent too. In our experiment, idempotent methods include HEAD, GET, OPTIONS, PUT, and DELETE.
\item[GET and POST:] HTML form element supports GET and POST methods and these methods are widely used in RPC services, hence we have categorized these methods separately.
\item[All:] This category accumulates all the seven methods listed in Table~\ref{tab:methods}, commonly used in RESTful services. It does not include all the methods listed in Table~\ref{tab:methodsum}.
\end{description}

\begin{table}[!t]
  \centering
  \caption{Categorized Method Support Distribution}
  \label{tab:methodgroup}
  \begin{tabular}{p{2.6cm} | >{\raggedleft\arraybackslash}p{3.4cm} >{\raggedleft\arraybackslash}p{2.2cm}}
    \hline
    \textbf{{\color{black}Category}} & \textbf{{\color{black}Supported (Count)}} & \textbf{{\color{black}Percentage}}\\
    \hline
    Safe & 22691 & 55.520\%\\
    Non-safe & 418 & 1.023\%\\
    Idempotent & 633 & 1.549\%\\
    GET and POST & 16497 & 40.365\%\\
    All & 418 & 1.023\%\\
    \hline
  \end{tabular}
\end{table}

\begin{table}[!t]
  \centering
  \caption{Interleaved Method Support Distribution}
  \label{tab:methodinter}
  \begin{tabular}{p{1.8cm} p{1.2cm} p{1.1cm} p{1.1cm} p{1.1cm} p{1.6cm} p{1.4cm} | >{\raggedleft\arraybackslash}p{0.8cm} >{\raggedleft\arraybackslash}p{1.2cm}}
    \hline
    \textbf{{\color{black}OPTIONS}} & \textbf{{\color{black}HEAD}} & \textbf{{\color{black}GET}} & \textbf{{\color{black}POST}} & \textbf{{\color{black}PUT}} & \textbf{{\color{black}DELETE}} & \textbf{{\color{black}PATCH}} & \textbf{{\color{black}Count}} & \textbf{{\color{black}\%}}\\
    \hline
    {\color{red}No} & {\color{red}No} & {\color{red}No} & {\color{red}No} & {\color{red}No} & {\color{red}No} & {\color{red}No} & {\color{black}17951} & {\color{red}43.922}\\[-0.25cm]
    {\color{red}No} & {\color{blue}Yes} & {\color{red}No} & {\color{red}No} & {\color{red}No} & {\color{red}No} & {\color{red}No} & {\color{black}4} & {\color{black}0.010}\\[-0.25cm]
    {\color{red}No} & {\color{blue}Yes} & {\color{red}No} & {\color{blue}Yes} & {\color{red}No} & {\color{red}No} & {\color{red}No} & {\color{black}2} & {\color{black}0.005}\\[-0.25cm]
    {\color{red}No} & {\color{blue}Yes} & {\color{blue}Yes} & {\color{red}No} & {\color{red}No} & {\color{red}No} & {\color{red}No} & {\color{black}99} & {\color{black}0.242}\\[-0.25cm]
    {\color{red}No} & {\color{blue}Yes} & {\color{blue}Yes} & {\color{blue}Yes} & {\color{red}No} & {\color{red}No} & {\color{red}No} & {\color{black}38} & {\color{black}0.093}\\[-0.25cm]
    {\color{red}No} & {\color{blue}Yes} & {\color{blue}Yes} & {\color{blue}Yes} & {\color{blue}Yes} & {\color{red}No} & {\color{red}No} & {\color{black}46} & {\color{black}0.113}\\[-0.25cm]
    {\color{red}No} & {\color{blue}Yes} & {\color{blue}Yes} & {\color{blue}Yes} & {\color{blue}Yes} & {\color{blue}Yes} & {\color{red}No} & {\color{black}4} & {\color{black}0.010}\\[-0.25cm]
    {\color{blue}Yes} & {\color{red}No} & {\color{red}No} & {\color{red}No} & {\color{red}No} & {\color{red}No} & {\color{red}No} & {\color{black}3} & {\color{black}0.007}\\[-0.25cm]
    {\color{blue}Yes} & {\color{red}No} & {\color{red}No} & {\color{red}No} & {\color{red}No} & {\color{blue}Yes} & {\color{red}No} & {\color{black}17} & {\color{black}0.042}\\[-0.25cm]
    {\color{blue}Yes} & {\color{red}No} & {\color{red}No} & {\color{red}No} & {\color{blue}Yes} & {\color{red}No} & {\color{red}No} & {\color{black}1} & {\color{black}0.002}\\[-0.25cm]
    {\color{blue}Yes} & {\color{blue}Yes} & {\color{red}No} & {\color{blue}Yes} & {\color{red}No} & {\color{red}No} & {\color{red}No} & {\color{black}14} & {\color{black}0.034}\\[-0.25cm]
    {\color{blue}Yes} & {\color{blue}Yes} & {\color{blue}Yes} & {\color{red}No} & {\color{red}No} & {\color{red}No} & {\color{red}No} & {\color{black}6264} & {\color{blue}15.327}\\[-0.25cm]
    {\color{blue}Yes} & {\color{blue}Yes} & {\color{blue}Yes} & {\color{red}No} & {\color{red}No} & {\color{blue}Yes} & {\color{red}No} & {\color{black}13} & {\color{black}0.032}\\[-0.25cm]
    {\color{blue}Yes} & {\color{blue}Yes} & {\color{blue}Yes} & {\color{red}No} & {\color{blue}Yes} & {\color{red}No} & {\color{red}No} & {\color{black}10} & {\color{black}0.024}\\[-0.25cm]
    {\color{blue}Yes} & {\color{blue}Yes} & {\color{blue}Yes} & {\color{red}No} & {\color{blue}Yes} & {\color{blue}Yes} & {\color{red}No} & {\color{black}11} & {\color{black}0.027}\\[-0.25cm]
    {\color{blue}Yes} & {\color{blue}Yes} & {\color{blue}Yes} & {\color{blue}Yes} & {\color{red}No} & {\color{red}No} & {\color{red}No} & {\color{black}15746} & {\color{blue}38.527}\\[-0.25cm]
    {\color{blue}Yes} & {\color{blue}Yes} & {\color{blue}Yes} & {\color{blue}Yes} & {\color{red}No} & {\color{blue}Yes} & {\color{red}No} & {\color{black}23} & {\color{black}0.056}\\[-0.25cm]
    {\color{blue}Yes} & {\color{blue}Yes} & {\color{blue}Yes} & {\color{blue}Yes} & {\color{blue}Yes} & {\color{red}No} & {\color{red}No} & {\color{black}2} & {\color{black}0.005}\\[-0.25cm]
    {\color{blue}Yes} & {\color{blue}Yes} & {\color{blue}Yes} & {\color{blue}Yes} & {\color{blue}Yes} & {\color{blue}Yes} & {\color{red}No} & {\color{black}204} & {\color{black}0.499}\\[-0.25cm]
    {\color{blue}Yes} & {\color{blue}Yes} & {\color{blue}Yes} & {\color{blue}Yes} & {\color{blue}Yes} & {\color{blue}Yes} & {\color{blue}Yes} & {\color{black}418} & {\color{red}1.023}\\
    \hline
  \end{tabular}
\end{table}

Table~\ref{tab:methodsum} provides quantities of individual methods, but it does not show how much those quantities overlap among various methods. Table~\ref{tab:methodinter} on the other hand represents an interleaved distribution of support of common HTTP methods. This table can be used to determine the quantity of support for any combination of common HTTP methods. If a combination is missing from the table, it has zero occurrences in our sample set. According to this table, 43.922\% live URIs either did not return an ``Allow'' header in the response to an OPTIONS request or they did not claim support for any of the seven common HTTP methods in the ``Allow'' header. 15.327\% live URIs claim support for OPTIONS, HEAD, and GET, but no other common methods. 38.527\% live URIs claim support for OPTIONS, HEAD, GET, and POST, but no other common methods. There are only 1.023\% URIs that claim support for all seven common HTTP methods in their ``Allow'' header in response to an OPTIONS request. All other combinations in Table~\ref{tab:methodinter} constitute only 1.201\% combined.

\begin{table}[!t]
  \centering
  \caption{Method Support Distribution across Web Server Software in \%}
  \label{tab:methodserver}
  \begin{tabular}{p{1.8cm} >{\raggedleft\arraybackslash}p{1.0cm} >{\raggedleft\arraybackslash}p{0.9cm} | >{\raggedleft\arraybackslash}p{1.0cm} >{\raggedleft\arraybackslash}p{1.2cm} >{\raggedleft\arraybackslash}p{1.1cm} >{\raggedleft\arraybackslash}p{1.1cm} >{\raggedleft\arraybackslash}p{1.0cm} >{\raggedleft\arraybackslash}p{1.0cm} >{\raggedleft\arraybackslash}p{1.0cm}}
    \hline
    \textbf{{\color{black}Server}} & \textbf{{\color{black}Count}} & \textbf{{\color{black}Allow}} & \textbf{{\color{black}OPT.}} & \textbf{{\color{black}HEAD}} & \textbf{{\color{black}GET}} & \textbf{{\color{black}POST}} & \textbf{{\color{black}PUT}} & \textbf{{\color{black}DEL.}} & \textbf{{\color{black}PAT.}}\\
    \hline
    Apache & 26071 & 14037 & 53.623 & 53.803 & 53.807 & 47.900 & 2.002 & 1.837 & 1.550\\
    IIS & 6371 & 5494 & 86.172 & 85.921 & 85.921 & 31.392 & 0.314 & 0.879 & 0.016\\
    Nginx & 3278 & 733 & 22.270 & 22.270 & 22.300 & 20.836 & 0.732 & 0.732 & 0.092\\
    Squeegit & 696 & 696 & 100.000 & 100.000 & 100.000 & 0.000 & 0.000 & 0.000 & 0.000\\
    ATS & 562 & 560 & 99.644 & 99.644 & 99.644 & 99.644 & 0.000 & 0.000 & 0.000\\
    Squid & 369 & 0 & 0.000 & 0.000 & 0.000 & 0.000 & 0.000 & 0.000 & 0.000\\
    YTS & 362 & 189 & 52.210 & 52.210 & 52.210 & 4.144 & 0.000 & 0.000 & 0.000\\
    AkamaiGHost & 342 & 0 & 0.000 & 0.000 & 0.000 & 0.000 & 0.000 & 0.000 & 0.000\\
    LiteSpeed & 300 & 0 & 0.000 & 0.000 & 0.000 & 0.000 & 0.000 & 0.000 & 0.000\\
    Lighttpd & 138 & 45 & 32.609 & 32.609 & 32.609 & 32.609 & 0.000 & 0.000 & 0.000\\
    Zeus & 139 & 55 & 0.000 & 39.568 & 39.568 & 0.000 & 0.000 & 0.000 & 0.000\\
    Unknown & 773 & 385 & 49.288 & 49.677 & 49.806 & 14.360 & 5.692 & 5.692 & 0.129\\
    Others & 1344 & 411 & 24.777 & 30.357 & 30.506 & 21.057 & 6.399 & 6.473 & 0.670\\
    \hline
  \end{tabular}
\end{table}

Table~\ref{tab:methodserver} shows the method support distribution in various web server software. Apache~\cite{apch} being the most popular web server, served 26,071 URIs in our sample set and returned ``Allow'' header in 14,037 responses. Over 53\% URIs served by Apache has claimed support for OPTIONS, HEAD, and GET methods, about 47\% claimed support for POST method, while only about 2\% or less supported PUT, DELETE, and PATCH methods. URIs using Microsoft IIS server~\cite{iis} claimed better support for OPTIONS, HEAD, and GET methods than those using Apache, but support for the rest of the methods was poorer than Apache. We have abbreviated names of OPTIONS, DELETE, and PATCH methods to OPT., DEL., and PAT. respectively to accommodate them in the table. It is worth noting that the values returned in the ``Server'' header were long strings that we have normalized to their common names. In our sample set 773 URIs that did not return ``Server'' header, about 385 URIs of those have returned ``Allow'' header, we have accumulated then under ``Unknown'' category. Any server that has less than 100 occurrences in our sample set was accumulated under ``Others'' category. Not all of the names in the Table~\ref{tab:methodserver} are first class web servers, for example, Squid is a proxy server, ATS (Apache Traffic Server) and AkamaiGHost are caching servers, and Squeegit is the name chosen by a free web hosting company called Tripod.com~\cite{tripod} for their web server. In our sample set, some server names were overwritten by the server control panel software. According to the HTTP specification, the ``Server'' header contains information about the origin server, hence proxy servers must not overwrite this header, they should use a ``Via'' header~\cite{rfc2616} instead.

\begin{table}[!t]
  \centering
  \caption{Method Support Distribution over Path Depths in \%}
  \label{tab:methoddepth}
  \begin{tabular}{p{1.8cm} >{\raggedleft\arraybackslash}p{1.0cm} >{\raggedleft\arraybackslash}p{0.9cm} | >{\raggedleft\arraybackslash}p{1.0cm} >{\raggedleft\arraybackslash}p{1.2cm} >{\raggedleft\arraybackslash}p{1.1cm} >{\raggedleft\arraybackslash}p{1.1cm} >{\raggedleft\arraybackslash}p{1.0cm} >{\raggedleft\arraybackslash}p{1.0cm} >{\raggedleft\arraybackslash}p{1.0cm}}
    \hline
    \textbf{{\color{black}Depth}} & \textbf{{\color{black}Count}} & \textbf{{\color{black}Allow}} & \textbf{{\color{black}OPT.}} & \textbf{{\color{black}HEAD}} & \textbf{{\color{black}GET}} & \textbf{{\color{black}POST}} & \textbf{{\color{black}PUT}} & \textbf{{\color{black}DEL.}} & \textbf{{\color{black}PAT.}}\\
    \hline
      0 & 27718 & 15302 & 54.863 & 55.058 & 55.123 & 39.895 & 1.490 & 1.465 & 0.978\\
      1 & 5141 & 3601 & 69.092 & 69.986 & 70.006 & 51.099 & 1.751 & 1.731 & 1.264\\
      2 & 4616 & 2325 & 49.588 & 50.347 & 50.368 & 34.879 & 2.643 & 2.708 & 1.040\\
      3 & 1689 & 1085 & 63.766 & 64.121 & 64.121 & 43.280 & 2.191 & 2.191 & 1.303\\
      4--12 & 1706 & 612 & 35.229 & 35.873 & 35.873 & 27.608 & 1.993 & 1.934 & 0.703\\
    \hline
  \end{tabular}
\end{table}

Table~\ref{tab:methoddepth} is similar to Table~\ref{tab:methodserver}, except it represents analysis of HTTP method support on URIs with different path depths as opposed to the server software used. It shows that the URIs with one path depth have over all better method support as compared with the URIs with zero path depth (root URIs). URIs with two path depth are dominating on PUT and DELETE method support while URIs with three path depth dominate on PATCH method support. There were not enough samples in higher order path depths (4--12), hence we have combined them together.

All results in Tables~\ref{tab:methodsum}, \ref{tab:methodgroup}, \ref{tab:methodinter}, \ref{tab:methodserver}, and \ref{tab:methoddepth} are based on the ``Allow'' header in the response of an OPTIONS request. We did not check to see if the URIs respond to the methods returned in the ``Allow'' header. Actual support may differ from what was claimed in the ``Allow'' header. For instance, all 40,870 live URIs returned \texttt{200 OK} response to a GET request, while only 22,899 (56.029\%) live URIs claimed support for GET method in the ``Allow'' header.

\section{Conclusion}

We have sampled 40,870 live URIs from DMOZ archived collection. We issued OPTIONS request against all the sampled URIs to collect their response code and response headers. Then we looked at the ``Allow'' response header to identify the supported HTTP methods. We found that 43.922\% live URIs either did not return an ``Allow'' header in the response to an OPTIONS request or they did not claim support for any of the seven common HTTP methods in the ``Allow'' header. 15.327\% live URIs claim support for OPTIONS, HEAD, and GET, but no other common methods. 38.527\% live URIs claim support for OPTIONS, HEAD, GET, and POST, but no other common methods. There are only 1.023\% URIs that claim support for all seven common HTTP methods in their ``Allow'' header in response to an OPTIONS request. We did not check to see if the URIs respond to the methods returned in the ``Allow'' header.

\bibliographystyle{plain}
\bibliography{httpoptions}

\end{document}